\documentclass[prl,aps,twocolumn,showpacs,floatfix,superscriptaddress]{revtex4-1}

\usepackage{graphicx}
\usepackage{dcolumn}
\usepackage{bm}
\usepackage{latexsym}
\usepackage{amsmath}
\usepackage{amssymb}
\usepackage[latin1]{inputenc}
\usepackage[usenames]{color}
\usepackage{wasysym}

\begin{document}

\title{Dielectric relaxation of water below the melting point: the effect of inner pressure.}

\author{Abril Angulo-Sherman and Hilda Mercado-Uribe}

\affiliation{
CINVESTAV-Monterrey, PIIT, Apodaca, Nuevo Le\'on 66600, Mexico}

\date{\today}

\begin{abstract}

 Despite water is the most studied substance in the Earth, it is not completely understood why its structural and dynamical properties 
 give rise to some anomalous behaviors. Interesting properties emerge when experiments at low temperatures and/or high pressures, are performed. Here we report dielectric measurements of cold water under constrained conditions, i.e. water that below the melting point can not freeze. The inner pressure shifts the $\alpha$ relaxation peak to similar frequencies as seen in ice Ih. Also, when we reach the triple point at 251 K, ice III seems to form. As far as we know, this via to obtain such crystalline phase has not been observed.
   
\end{abstract}


\maketitle

Since the seminal works of Canton {\it et al} in the eighteen century \cite{Canton1761,Canton1764},  water has been extensively studied to understand its structure and dynamical behavior. Due to its small molecular size, its polarity, and most importantly, the interaction through hydrogen bonds, water exhibits many special properties. Most authors agree to call such properties as anomalous \cite{Chaplin2001,Finney2004}, and a sophisticated phase diagram is an evidence of it \cite{Chaplinsite}. Water has eleven stable triple points and a remarkable versatility in crystalline structure forming, like no other known substance \cite{Malenkov2009,Finney2004}. 
As the temperature decreases at atmospheric or hyperbaric pressures, water molecules rearrange themselves to occupy either less volume or more, changing their length and angle of the hydrogen bonds. Different paths exists to form the crystalline structures. 

Below $273$ K, water can exist in a liquid state in precarious equilibrium. When it is cooled around $10^6$ K/s, freezing can be avoided and water remains in a supercooled state, until it turns into glass \cite{Angell1983,Mishima1998,Debenedetti2003}. On the other hand, crystalline phases are usually obtained by pressurization produced by a piston and gasses at low temperatures \cite{Chan1965,Wilson1965,Gough1970,Johari1976,Johari1981,Andersson2007,Loerting2008}. In particular,
 ice III is generally obtained by three different ways: cooling the liquid at about 300 MPa, warming ice II at about 300 MPa, or decompressing ice V at about 247 K \cite{Wilson1965}. 

Dielectric spectroscopy has proven to be a powerful technique to investigate the molecular behavior of water. Particularly because the relaxation times provide information about the cooperative processes, proton mobility, viscosity of the medium and, in general, structural  properties of the molecular network \cite{Johari1981,Minoguchi2004,Swenson2010}. This is the technique we use in work reported in the present Letter, where dielectric measurements of water at low temperatures and constant volume are performed. At these conditions, as we decrease the temperature the pressure diverges. This internal pressure shifts the Debye relaxation to lower frequencies, the very same than ice Ih. At temperatures below 251 K, the relaxation times deviate and slightly increase, possibly due to the formation of a thin shell of ice III and Ih.

Our experimental measurements were carried out in a tight and long cylindrical capacitor able to endure inner pressures as high as 200 MPa. The results here reported were obtained using pure water (Mili-Q, $18.2$ M$\Omega$cm). The setup is similar to the one employed in a previous paper \cite {Abril2011}, with the following modifications: the cylindrical stainless steel capacitor is 15 cm long, surrounded by an aluminium block with channels where dynalene flows through in order to regulate the temperature. Two different samples were studied. First, the capacitor was totally filled with 6.87 ml of water and sealed with two rigid nylon caps that are screwed to the outer electrode. Second, $90 \%$ of the above volume was poured inside the capacitor, cool it down to freeze and then closed. The dielectric properties, $\epsilon^\prime$ and $\tan \delta$ ($\tan \delta = \epsilon^{\prime\prime}/\epsilon^\prime$), were measured using a LCR meter (HP 4284A) in the interval of frequency $10^2$ - 3x$10^5$ Hz. The temperature of the device was controlled by a PolySciences circulating bath from 277 to 239 K in steps of 2 K, with a thermal stability of $\pm 0.1^\circ$  K. After each temperature is reached, we waited approximately 60 min to get thermal equilibrium in the whole capacitor.

\begin{figure}[ht!]
\begin{center}
\includegraphics[scale=0.86]{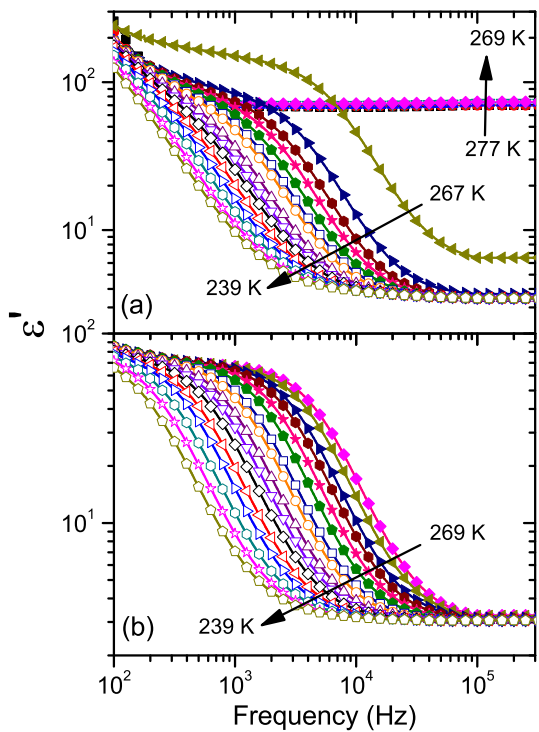}
\caption{\small (Color online) Relative permittivity as a function of frequency from 277 to 239 K in steps of 2 K for (a) water and (b) ice Ih. The symbols are the experimental data and the lines are the fits as described in the text.}
\label{fig1}
\end{center}
\end{figure}

\begin{figure}[ht!]
\begin{center}
\includegraphics[scale=0.86]{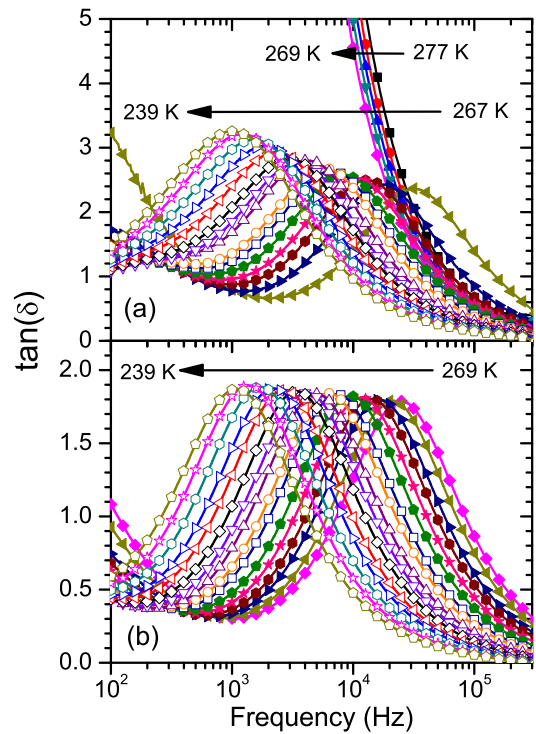}
\caption{\small (Color online) Dielectric loss tangent as a function of frequency from 277 to 239 K in steps of 2 K for (a) water and (b) ice Ih. The symbols are the experimental data and the lines are the fits as described in the text.}
\label{fig2}
\end{center}
\end{figure}
	
Figure \ref{fig1} shows the relative permittivity as a function of frequency for water (a) and ice Ih (b) at different temperatures. The main aim of this work was to study the behavior of the first, but we show the results of ice Ih (a phase well studied before \cite{Chan1965,Kawada1979,Johari1976,Johari1981}) only for comparative purposes. It can be observed in (a) the usual dielectric response of water from 277 to 269 K, composed by a region of a very high relative permittivity at the lowest frequencies, caused by a Maxwell-Wagner-Sillars (MWS) polarization, and a second region where the dipole response gives a flat contribution. From 267 to 239 K the Debye relaxation moves to lower frequencies as the temperature decreases. Note that an intriguing response occurs at 267 K: the whole spectrum separates notoriously from the rest of the curves (this result will be later discussed).

Although the dielectric loss signal is normally given by the imaginary part of complex dielectric function, we prefer to plot the $\tan \delta$ in order to highlight the details of the main relaxation peak, see Fig. \ref{fig2}. Again, there is a clear distinction between the response obtained at 267 K from those found at lower temperatures. Moreover, while the maximum amplitudes of ice Ih are quite the same regardless the temperature, for water they increase as the temperature decreases.

\begin{figure}[ht!]
\begin{center}
\includegraphics[scale=0.93]{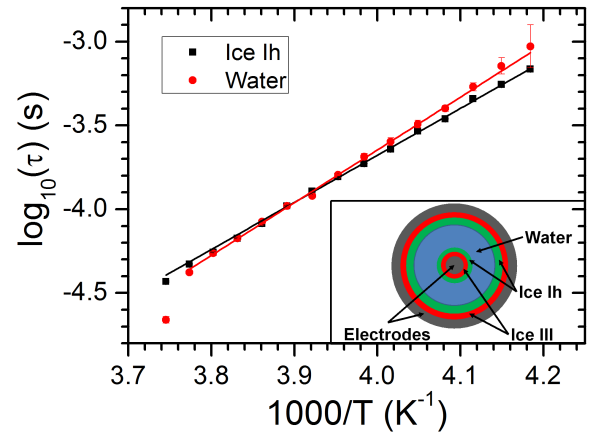}
\caption{\small (Color online) Dielectric relaxation times as a function of inverse temperature for water (red circles) and ice Ih (black squares). In the inset it is shown the cross section of the capacitor with water surrounded by thin shells of ice Ih and ice III.}
\label{fig3}
\end{center}
\end{figure}

\begin{figure}[ht!]
\begin{center}
\includegraphics[scale=0.83]{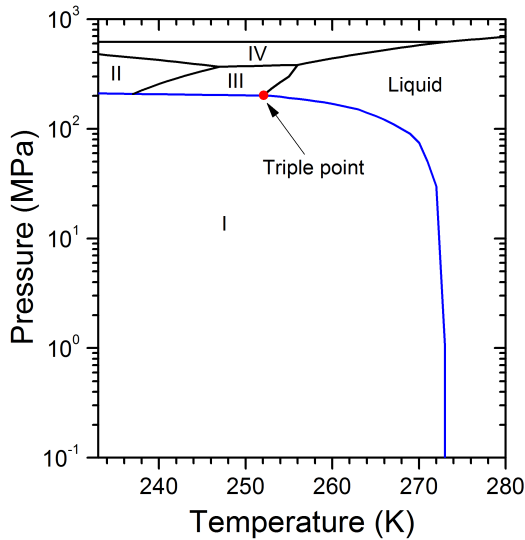}
\caption{\small (Color online) A section of the phase diagram of water in the temperature-pressure range used in this work \cite{Chaplinsite}.}
\label{fig4}
\end{center}
\end{figure}

The dielectric spectra of Figs. \ref{fig1} and \ref{fig2} were fitted to an expression that takes into account the $\alpha$ relaxation, the MWS electrode polarization and the dc-conductivity \cite{Havriliak1967,Richert2010}. 




Figure \ref{fig3} shows the dielectric relaxation times obtained from such fits versus the reciprocal temperature for water and ice Ih. The relaxation times for water are similar to the ones reported for ice Ih \cite{Gough1970,Johari1981,Bruni1993} but they start to separate at  251 K.  In order to understand this behavior, let us see Fig. \ref{fig4}, which depicts the water phase diagram for pressures below $10^3$ MPa \cite{Chaplinsite} in the temperature range of this work. It is evident that, as the liquid is cooled below 273 K, the thermodynamic path follows the coexistence line where the pressure climbs to very high values. Indeed, according to the Clausius-Clapeyron equation $ \frac{dP}{dT} = \frac{\Delta H}{T \Delta V} $  (where P, T, H and V are the pressure, temperature, enthalpy and volume, respectively), the pressure augments considerably due to the fact the volume is not allowed to increase. This inner pressure acts against the electrodes and when it reaches approximately $210$ MPa (at 251 K), a shell of ice III starts to nucleate on the electrode surfaces. Furthermore, due to the fact ice III is denser than water, as soon as it appears, it leaves space for a thin layer of ice Ih to form. In the inset of Fig. \ref{fig3} we show a schematic of this situation. Obviously, liquid water must necessarily remain in the interior because neither ice Ih nor ice III can completely fill its volume. It is enthralling to think that ice III can not fill the volume because, in the absence of external pressure, who else would exert the pressure it needs to exist?

The relaxation times shown in Fig. \ref{fig3} have an Arrhenius behavior, but it is unexpected that they are of the same order. One could think that this is natural due to the fact both phases coexist. However, it is not at all easy to see how the relaxation times of water are around 8 orders of magnitude greater than normal water \cite{Agmon1996,Buchner1999,Okada1999,Bertolini1982}. It is plausible to explain such outcome if we draw up on previous works \cite{Ludemann1987,Kaatze2002}. There, it is stated that when water is compressed at very high pressures the molecules increase the number of near neighbours and the tetrahedrality force field of the network is broken. Consequently, the energy landscape is less partitioned and molecules rotate easily. Under these conditions the applied electric field can not align the water dipoles as easily unless the frequency of the field is low. Therefore the relaxation times are large. In other words, in ice Ih the dipoles are fixed, in water under pressure are free. It is fascinating to find that the relaxation times are similar.

Last but not least, we now go back to Fig. \ref{fig1} where the curve corresponding to 267 K is puzzling. As we previously mentioned, the high pressure increases molecular rotation. This event requires that at least two hydrogen bonds break \cite{Bagchi2005}. Coincidentally, such reorganization of the hydrogen bonds propel Grotthus mechanism, which is related to a fast transport of protons \cite{Agmon1995}. Thus, we believe this is the only way to explain a very high value of $\epsilon^\prime$ (note that at frequencies where the electric polarization is normally not existing, $10^3-10^4$ Hz, $\epsilon^\prime$ is much greater than 100). As far as we know, this is the first time a dielectric signature of Grotthus mechanism has been observed.


Overall, we have found a subtle but clear signal of  ice III formed when water is cooled below its melting point in a constrained volume. With it, we believe ice Ih is also formed. We must emphasize that this water-ice III-ice Ih coexistence occurs at complete  equilibrium, in contrast to the metastable state of supercooled water. We show that the inner pressure shifts the $\alpha$ peak of normal water to low frequencies (kHz). Finally, we found an intriguing dielectric signal at 267 K probably caused by a fast ionic transport. Our results may have important implications in the understanding of water dynamics under pressure.


\section{ACKNOWLEDGMENTS}

We thank C. Ruiz for encouragement and critical discussions. This work has benefited from the Grant 154202 given by CONACyT, Mexico. A. A. S. wishes to acknowledge a scholarship from CONACyT.


\end{document}